\documentclass[submission,copyright,creativecommons]{eptcs}
\providecommand{\event}{WACA 2025} 
\providecommand{\thisvolume}[1]{this volume of EPTCS, Open Publishing Association}

\usepackage{iftex}
\usepackage{graphicx}
\graphicspath{ {./images/} }
\usepackage[toc,page]{appendix}

\usepackage{wrapfig}

\usepackage{tabulary}
\usepackage{booktabs}             
\usepackage{calc}                 

\newlength{\tabwidthi}
\newlength{\tabwidthii}
\newlength{\tabwidthiii}

\newlength{\tempwidth}

\newcommand{\tabsize}{\footnotesize}

\usepackage{soul}
\soulregister{\bf}{0}
\soulregister{\em}{0}
\soulregister{\cf}{7}
\soulregister{\eg}{7}
\soulregister{\Eg}{7}
\soulregister{\etc}{7}
\soulregister{\ie}{7}
\soulregister{\resp}{7}
\soulregister{\wrt}{7}
\soulregister{\code}{7}
\soulregister{\cite}{7}
\soulregister{\pageref}{7}
\soulregister{\ref}{7}
\soulregister{\chap}{7}
\soulregister{\defn}{7}
\soulregister{\eq}{7}
\soulregister{\eqs}{7}
\soulregister{\fig}{7}
\soulregister{\lem}{7}
\soulregister{\prop}{7}
\soulregister{\secn}{7}
\soulregister{\thm}{7}
\soulregister{\assm}{7}

\usepackage{listings}
\usepackage{xcolor}  

\usepackage{hyperref}
\hypersetup{
    colorlinks=true,
    linkcolor=blue,      
    urlcolor=blue,
}

\definecolor{codegreen}{rgb}{0,0.6,0}
\definecolor{codegray}{rgb}{0.5,0.5,0.5}
\definecolor{codepurple}{rgb}{0.58,0,0.82}
\definecolor{backcolour}{rgb}{0.95,0.95,0.92}

\lstdefinestyle{mystyle}{
    backgroundcolor=\color{backcolour},   
    commentstyle=\color{codegreen},
    keywordstyle=\color{magenta},
    numberstyle=\tiny\color{codegray},
    stringstyle=\color{codepurple},
    basicstyle=\ttfamily\footnotesize,
    breakatwhitespace=false,         
    breaklines=true,                 
    captionpos=b,                    
    keepspaces=true,                 
    numbers=left,                    
    numbersep=5pt,                  
    showspaces=false,                
    showstringspaces=false,
    showtabs=false,                  
    tabsize=2,
    frame=single
}

\ifpdf
  \usepackage{underscore}         
  \usepackage[T1]{fontenc} 
  \usepackage[utf8]{inputenc} 
\else
  \usepackage{breakurl}           
\fi

\title{AdaptiFlow: An Extensible Framework for\\Event-Driven Autonomy in Cloud Microservices}
\author{Brice Arléon Zemtsop Ndadji
\institute{Univ. Lille, CNRS, Inria, Centrale Lille, UMR 9189 CRIStAL, F-59000 Lille, France}
\email{brcie-arleon.zemtsop-ndadji@inria.fr}
\and
Simon Bliudze
\institute{Univ. Lille, Inria, CNRS, Centrale Lille, UMR 9189 CRIStAL, F-59000 Lille, France}
\email{simon.bliudze@inria.fr}
\and
Clément Quinton
\institute{Univ. Lille, CNRS, Inria, Centrale Lille, UMR 9189 CRIStAL, F-59000 Lille, France}
\email{clement.quinton@inria.fr}
}
\def\titlerunning{AdaptiFlow: An Extensible Framework for Event-Driven Autonomy in Cloud Microservices}
\def\authorrunning{B.A. Zemtsop Ndadji, S. Bliudze \& C. Quinton}

\newcommand{\mdash}[1][]{#1---#1}

\newcommand{\eg}[1][\@ ]{e.g.#1}
\newcommand{\etc}[1][]{etc.#1}
\newcommand{\cf}[1][~]{cf.#1}
\newcommand{\wrt}[1][\@ ]{w.r.t.#1}

\begin{document}
\maketitle

\begin{abstract}
Modern cloud architectures demand self-adaptive capabilities to manage dynamic operational conditions. Yet, existing solutions often impose centralized control models ill-suited to microservices' decentralized nature. This paper presents \textbf{AdaptiFlow}, a framework that leverages well-established principles of autonomous computing to provide abstraction layers focused on the \textbf{Monitor} and \textbf{Execute} phases of the MAPE-K loop. 
By decoupling metrics collection and action execution from adaptation logic, AdaptiFlow enables microservices to evolve into autonomous elements through standardized interfaces, preserving their architectural independence while enabling system-wide adaptability. The framework introduces: (1) \textbf{Metrics Collectors} for unified infrastructure/business metric gathering, (2) \textbf{Adaptation Actions} as declarative actuators for runtime adjustments, and (3) a lightweight \textbf{Event-Driven} and rule-based mechanism for adaptation logic specification. Validation through the enhanced \textbf{Adaptable TeaStore} benchmark 
demonstrates practical implementation of three adaptation scenarios targeting three levels of autonomy\mdash self-healing (database recovery), self-protection (DDoS mitigation), and self-optimization (traffic management)\mdash with minimal code modification per service. Key innovations include a workflow for service instrumentation and evidence that decentralized adaptation can emerge from localized decisions without global coordination. The work bridges autonomic computing theory with cloud-native practice, providing both a conceptual framework and concrete tools for building resilient distributed systems. Future work includes integration with formal coordination models and application of adaptation techniques relying on AI agents for proactive adaptation 
to address complex adaptation scenarios.

\end{abstract}


\section{Introduction}  
Modern cloud architectures face growing complexity due to their distributed nature, necessitating systems that autonomously adapt to dynamic conditions. The MAPE-K loop \cite{kephart2003vision} (Monitor-Analyze-Plan-Execute-Knowledge) has long served as the foundation for self-adaptive systems, traditionally implemented as a centralized, reactive, and sequential loop for executing adaptations \cite{sanwouo2025breaking}. However, the decentralized nature of microservice-based applications demands a paradigm shift toward separating functional and adaptation concerns. Drawing inspiration from hardware abstraction layers (HALs) in operating systems, this paper introduces an abstraction layer focused on the \textbf{Monitor} and \textbf{Execute} phases of MAPE-K, enabling microservices to expose observable data and to accept control commands without invasive code changes. As envisioned in the seminal paper by Kephart and Chess~\cite{kephart2003vision}, this approach ensures "the managed element can be adapted to enable the autonomic manager to monitor and control it," bridging the gap between non-adaptive systems and self-managing elements.  

We present \textbf{AdaptiFlow}, a framework providing standardized interfaces to instrument microservices for self-adaptation.
AdaptiFlow’s architecture is designed to address the unique challenges of self-adaptive systems in cloud-native environments, where decentralization, scalability, and context-awareness are critical. Self-adaptive systems are capable of dynamically altering their structure and behavior during runtime by continuously evaluating their environment, internal state, and operational goals \cite{metzger2024realizing}. For instance, consider a microservice that responds to a sudden workload increase by disabling non-essential functionalities. An online retailer, for example, might temporarily deactivate its computationally intensive recommendation engine during peak traffic periods. Such adaptations enable the system to preserve Key Performance Indicators (KPIs, \eg latency, CPU, and memory usage) despite fluctuating demands. To build a self-adaptive system, developers must encode logic that defines the data to observe (\eg performance metrics), the events (\eg traffic increase, service unavailable, DDoS Attack), the conditions triggering adaptation, and the specific mechanisms for executing those changes.
AdaptiFlow’s core abstraction layer comprises:  
\begin{itemize}  
    \item \textbf{Metrics Collectors}: Unified APIs to gather infrastructure (CPU, latency) and business-level metrics (cache hits, transaction rates).  
    \item \textbf{Adaptation Actions}: Declarative interfaces to define infrastructure-level (\eg scaling) and busi\-ness-level (\eg feature toggling) actuators.  
\end{itemize}  

By decoupling monitoring and execution from adaptation logic, AdaptiFlow enables diverse strategies for analysis and planning. Developers can implement:  
\begin{itemize}  
    \item \textbf{Internal Logic}: Rule-based adaptations (\eg threshold-driven events) embedded directly into services.  
    \item \textbf{Exogenous Logic}: External agents or planners (\eg AI agents) leveraging AdaptiFlow’s interfaces to collect metrics and execute actions.  
\end{itemize}  

To validate this approach, AdaptiFlow incorporates event-driven concepts that enable rule-based adaptation logic. The framework provides \textit{Event Observation} mechanisms to detect threshold-based (\eg CPU > 80\%) or custom events (\eg service degradation patterns) through periodic polling or on-demand triggers. These events act as semantic bridges between raw metrics and actionable adaptations, allowing developers to declaratively specify \textit{when} and \textit{how} the system should respond to changing conditions. 

We validate AdaptiFlow on the \textit{Adaptable TeaStore Benchmark}~\cite{bliudze2024adaptable,bliudze2025impl}. Adaptable TeaStore is an extension of the TeaStore microservices benchmark~\cite{von2018teastore} supporting autonomous behaviour and specifying adaptation scenarios~\cite{bliudze2024adaptable}. An implementation of the benchmark is available~\cite{bliudze2025impl} for evaluating different approaches to adaptation. We apply AdaptiFlow to that implementation to realise the adaptation scenarios specified in~\cite{bliudze2024adaptable}. Our experiments demonstrate AdaptiFlow’s practicality through three implemented adaptation scenarios targeting three levels of autonomy: self-healing (service recovery), self-protection (DDoS mitigation), and self-optimization (traffic management) with minimal code changes.   

\paragraph{Key Contributions:}
\begin{itemize}  
    \item Design and validation of an abstraction layer for the Monitoring and Execution phases of the MAPE-K loop, transforming non-adaptive services into autonomic elements
    \item A workflow to instrument microservices with metric collectors and with actuators
    \item Empirical validation through the \emph{Adaptable TeaStore}
\end{itemize}  

The rest of the paper is structured as follows. Section~\ref{secn:context} presents the state of the art, Section~\ref{secn:fw-design} outlines the framework design. Section~\ref{secn:casestudy} details experimental results with the Adaptable TeaStore case study. Section~\ref{secn:conclusion} concludes the paper.

\section{State of the art}  
\label{secn:context} 

\settowidth{\tempwidth}{Figure 1: The MAPE-K loop}
\begin{wrapfigure}[13]{r}{1.2\tempwidth}
\vspace{-\baselineskip}
\centering  
\includegraphics[width=0.25\textwidth]{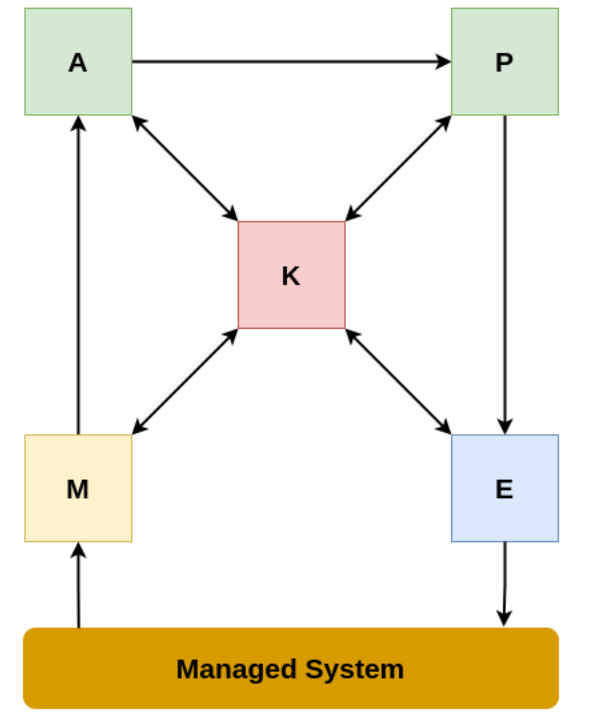}  
\caption{The MAPE-K loop}
\label{fig:mape-k-loop}  
\end{wrapfigure} 

Modern computing systems are evolving toward autonomic paradigms, where components manage their behavior autonomously to reduce human intervention. Inspired by biological systems, the vision of autonomic computing \cite{kephart2003vision} envisions distributed networks of \textit{autonomic elements}, self-managing entities that deliver services while adhering to predefined policies. Each element consists of a \textit{managed resource} (\eg a database, CPU, or microservice) and an \textit{autonomic manager} that monitors and controls it. Over time, the distinction between manager and managed resource may dissolve \cite{kephart2003vision}, yielding fully integrated elements capable of independent decision-making while collaborating through decentralized interactions.

More broadly, the evolution of self-adaptive systems, as chronicled by Weyns \cite{Weyns21}, can be understood through seven successive waves of research focus. These waves encompass: (1) \emph{Automating Tasks} with external managers \cite{Oreizy1999, kephart2003vision, dobson2006}, (2) \emph{Architecture-based Adaptation} for separation of concerns \cite{garlan2004, kramer2007, weyns2012}, (3) the use of \emph{Runtime Models} for reasoning \cite{blair2009, morin2009, vogel2014}, (4) \emph{Requirements-driven} design \cite{whittle2009, souza2013, iftikhar2014}, (5) providing \emph{Guarantees under Uncertainty} \cite{calinescu2010, moreno2015, calescu2017}, (6) \emph{Control-based} adaptation for theoretical guarantees \cite{hellerstein2004feedback, filieri2014automated, shevtsov2016keep, maggio2017automated}, and (7) \emph{Learning from Experience} with machine learning \cite{epifani2009model, elkhodary2010fusion, jamshidi2016fuzzy, quin2019efficient}.
This historical context frames the fundamental paradigms that influence the design of AdaptiFlow. Our work embraces the push toward separation of concerns, decentralization, and requirements-driven design. We argue that to realize these goals in modern cloud-native environments, an abstraction layer for adaptation logic analogous to a hardware abstraction layer is needed. AdaptiFlow aims to provide such a layer, built upon the established principles of the MAPE-K control loop \cite{kephart2003vision}, and rule-based adaptation techniques \cite{salatino2016mastering}. These concepts are essential for managing complex distributed systems, from early monolithic architectures to modern cloud-native environments. Figure~\ref{fig:mape-k-loop} provides a visual summary of the MAPE-K reference model that underpins much contemporary adaptation research.

The Monitor-Analyze-Plan-Execute over a shared Knowledge (MAPE-K) loop (Fig.~\ref{fig:mape-k-loop}), introduced over two decades ago by Kephart et al.~\cite{kephart2003vision}, remains a cornerstone of self-adaptive systems. It structures adaptation into four phases:  
\begin{itemize}  
    \item \textbf{M}onitor: Observes the system’s internal state and environment (\eg CPU usage, request latency).  
    \item \textbf{A}nalyze: Evaluates observations to identify adaptation needs (\eg detecting service failures).  
    \item \textbf{P}lan: Generates strategies to achieve desired states (\eg scaling instances, enabling circuit breakers).  
    \item \textbf{E}xecute: Implements selected adaptation actions.  
\end{itemize}  
A shared \textbf{K}nowledge (K) repository stores contextual data (\eg metrics, policies) to support decision-making across phases. Originally applied to monolithic systems, MAPE-K has since been adapted for cloud computing \cite{maurer2011revealing, karol2024self, metzger2024realizing}, IoT \cite{oh2022analysis, riegler2023distributed}, and cybersecurity \cite{jahan2020mape, jamshidi2024enhancing, stadler2024cyber}.

Rule-based systems like Drools \cite{salatino2016mastering} can be used to operationalize the \textbf{Analyze} and \textbf{Plan} phases of MAPE-K by codifying adaptation logic into declarative condition-action pairs. A rule’s structure "when \textit{condition}, then \textit{action}" enables developers to declaratively specify behaviors (\eg "when CPU > 80\%, then scale instances"). Unlike traditional code, rules are modular, auditable, and dynamically updatable, making them ideal for scenarios requiring rapid adjustments (\eg fraud detection, load management). For example, Drools filters data through conditions and triggers actions when matches occur. This approach bridges technical and business requirements, allowing non-developers to contribute to adaptation policies.  

Complementary to centralized frameworks, Gru \cite{florio2016gru} introduces a decentralized autonomic approach for microservices based on agents. Each \textit{Gru-Agent} embeds a local MAPE loop and independently manages a set of Docker containers on a node. Decisions such as scaling or migrating microservices are taken based on local and partial neighborhood state information, improving fault tolerance. Gru’s autonomy layer wraps microservices without modifying their internals. This design shares goals with AdaptiFlow (flexibility, separation of concerns), but Gru operates closer to the infrastructure level, making it a valuable complement for decentralized adaptation logic.

Recent contributions explore increasingly specialized mechanisms to improve system resilience. For example, Sedghpour et al.~\cite{sedghpour2021service} investigate circuit-breaking strategies embedded directly into the service mesh layer. Leveraging control theory, their adaptive controller dynamically adjusts queue limits to maintain response time guarantees under stress conditions.

While frameworks such as Drools and agent-based solutions like Gru have significantly simplified the expression of adaptation logic, we take on the challenge of providing a lightweight solution in the form of a reusable abstraction layer that exposes standard interfaces for both service monitoring and adaptation execution—spanning infrastructure-level and application-level concerns. This abstraction explicitly focuses on separating functional responsibilities from adaptation logic in cloud microservice systems. Unlike traditional frameworks, our approach does not impose a specific MAPE control pattern; rather, it grants developers the flexibility to adopt the integration pattern that best fits their needs. Moreover, it facilitates the construction of benchmarks of self-adaptive microservices on which adaptation approaches such as Sedghpour et al.~\cite{sedghpour2021service} can be implemented and compared.

\section{Framework Design}
\label{secn:fw-design}

AdaptiFlow's architecture envisages autonomous management for cloud-native systems through a decomposition of the MAPE-K loop. Our design introduces a modular approach based on two key principles: (1) standardization of observation and actuation interfaces, and (2) flexible integration of various adaptation strategies. This separation allows microservices to retain their architectural independence while participating in system-wide adaptation models. Figure~\ref{fig:arch-overview} provides a visual illustration of the AdaptiFlow specification within autonomic managers. This illustration showcases AdaptiFlow's flexibility, deployed here in a decentralized architecture where each autonomic element manages its own adaptation logic. The same abstraction layer can alternatively support a centralized coordination model, as demonstrated by the database recovery scenario in Section~\ref{subsec:self-healing}.

\subsection*{Architectural Overview}  
AdaptiFlow provides a set of directives to specify the adaptation logic. The framework operationalizes the MAPE-K loop by focusing on two phases \textbf{Monitor} and \textbf{Execute} tailored for decentralized cloud environments. We intentionally omit the \textbf{Analyze} and \textbf{Plan} phases since our goal is to allow the separation of concerns. However, in order to allow closing the loop, we do provide the mechanism for triggering actions when a corresponding condition is satisfied using an event-driven methodology. This is a lightweight implementation of the analysis and planning phases, which originally involve assessing the microservices execution context, identifying the system's adaptation needs, and finally generating adaptation strategies in order to achieve the adaptation objective. Since our solution is provided in the form of a Java library, it allows for the specification of arbitrary custom adaptation strategies. However, it intentionally does not provide any dedicated syntax or abstraction (a domain-specific language for adaptation) for that purpose. These are left for separate future work.

\begin{figure}[p]  
\centering  
\includegraphics[width=0.9\textwidth]{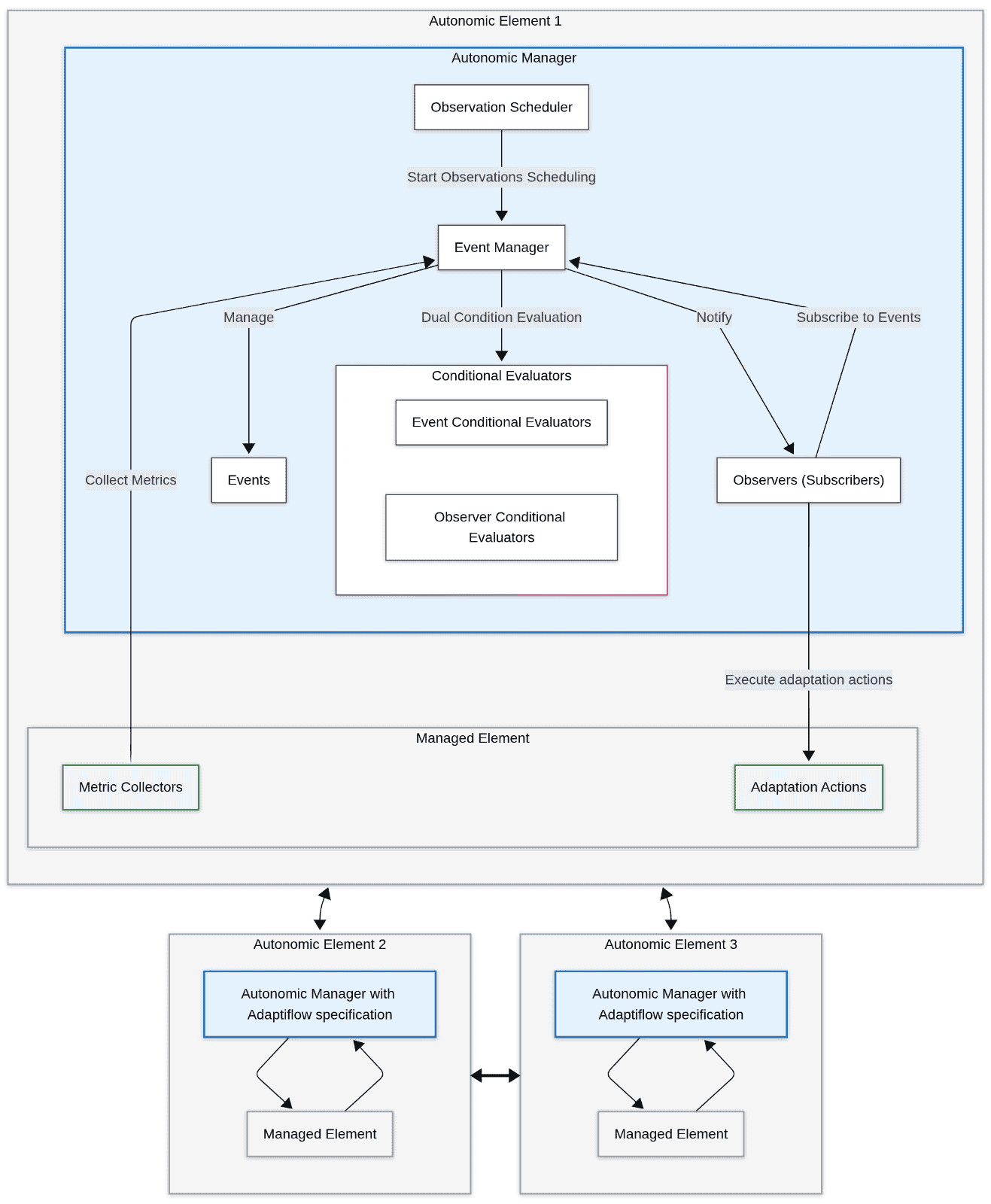}  
\caption{AdaptiFlow specification within autonomic managers of a decentralized architecture, emphasizing modular components for metrics collection, adaptation action execution, and event handling.}  
\label{fig:arch-overview}  
\end{figure}

In the \textbf{Monitor} phase, metrics collectors continuously gather infrastructure (\eg CPU, memory) and business-level (\eg request rates, service states) data, providing real-time insights into system state. For example, a \texttt{LatencyCollector} might track API response times, while a \texttt{Re\-source\-Usage\-Collector} monitors cloud resource availability. 

The \textbf{Execute} phase then invokes predefined or custom adaptation actions, such as scaling services via Kubernetes (\texttt{ScaleService}) or disabling non-critical features like recommendations (\textit{EnableFallbackAlgorithm}). AdaptiFlow uses a decentralized decision-making through event subscriptions: services subscribe to relevant events (\eg an auth service subscribing to \texttt{DDoSAttackEvent}).

The \textbf{Analyze} phase employs conditional evaluators to interpret the data collected, detecting events such as \texttt{HighWorkloadEvent} (request rate > 1,000/s OR CPU > 80\%) or \texttt{Service\-Failure\-Event} (health check timeout). These evaluators apply threshold-based rules (\eg CPU > 80\%) or custom logic (\eg apply more complex tests to the given metric) to determine adaptation triggers. 

The shared \textbf{Knowledge} (K) is derived dynamically from distributed metrics rather than a static repository. This approach offers three key advantages: (1) \textit{Decentralized control}, enabling parallel adaptations; (2) \textit{Transparency}, as rules and events are explicitly defined and auditable; and (3) \textit{Flexibility}, supporting hybrid adaptations that combine infrastructure adjustments (scaling) with business logic changes (feature toggling). AdaptiFlow distributes adaptation logic across microservices, aligning with their autonomous nature. This design choice enables granular, context-driven adaptations.

To realize this architecture, AdaptiFlow decomposes the adaptation process into six modular components (Fig.~\ref{fig:arch-overview}), each addressing a distinct aspect of self-adaptation. The (Fig.~\ref{fig:sequence-diagram}) describes the interactions between those components: 

\begin{figure}[t]
\centering
\includegraphics[width=1\textwidth]{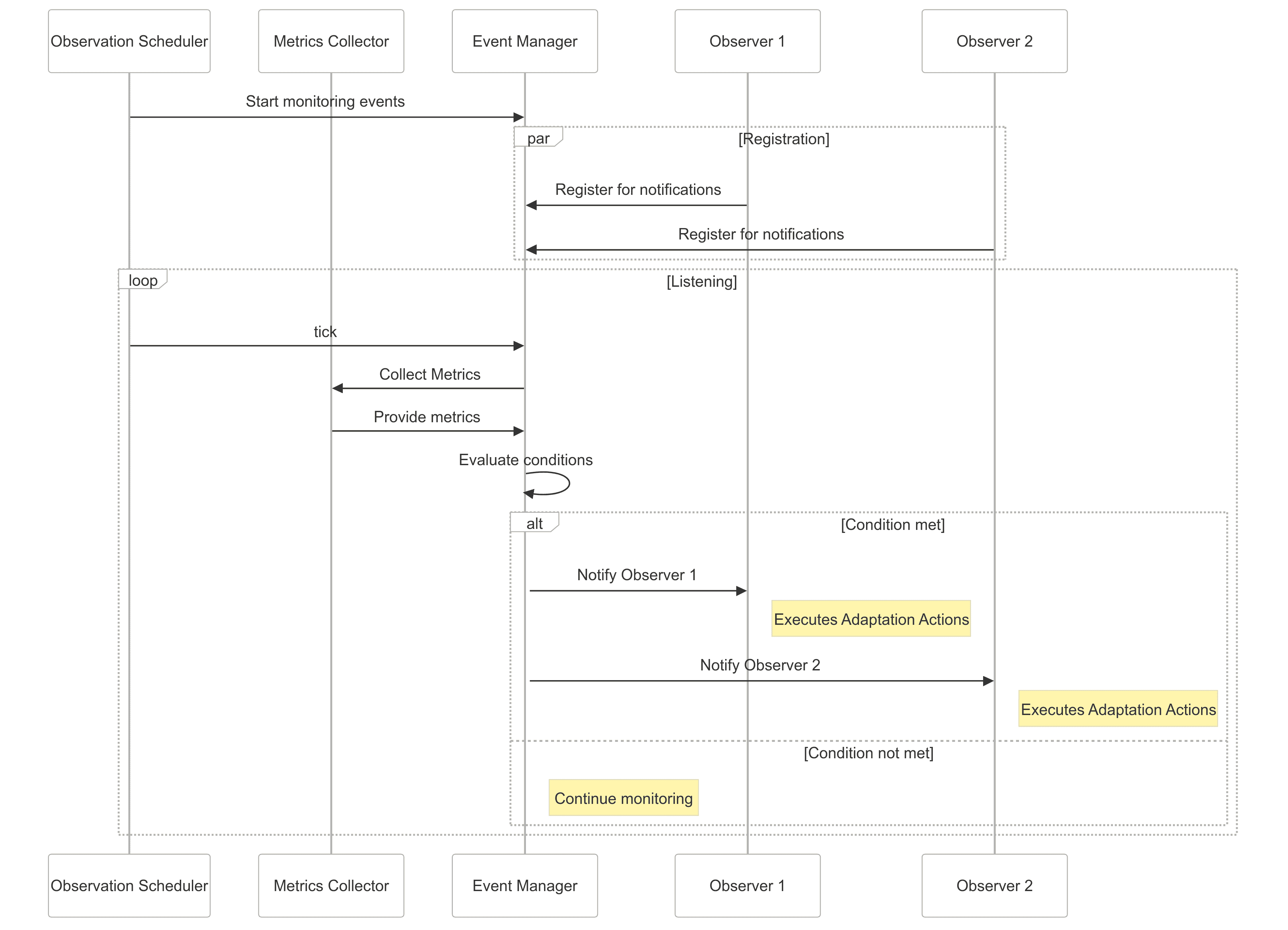}
\caption{Sequence diagram illustrating runtime interactions between AdaptiFlow components}
\label{fig:sequence-diagram}
\end{figure}

\begin{itemize}  
    \item \textbf{Metrics Collectors} (Perception of the Environment): gather infrastructure and business-level metrics (\eg CPU usage, API latency) to establish a real-time view of the context.  
    
    \item \textbf{Adaptation Actions} (Executing Change): translate decisions into runtime adjustments, ranging from infrastructure operations (\eg scaling) to business logic modifications (\eg enabling fallback algorithms).  
    
    \item \textbf{Event Management}: the events are specified using conditional evaluators, then services or service components subscribe to them and the event observation starts. When the conditions specified in the event conditional evaluator are met, the event is triggered and the event subscribers are notified.
    \begin{itemize} 
        \item \textbf{Conditional Evaluators} (Adaptive Decision-Making): apply threshold-based or custom logic (\eg "IF cache hit ratio < 20\% AND peak hour") to determine when events should trigger and also when subscribers should be notified to execute adaptations. Conditional Evaluators are part of the specification of both events and event subscribers.
    
        \item \textbf{Event Specification} (Contextual Awareness): defines the structure of an event by the use of conditional evaluators on the collected metrics (\eg \texttt{ResourceExhaustionEvent} = CPU > 90\% \&\& \texttt{FreeDisk} < 10\%). Developers implicitly specify event types (threshold-based or custom) and their triggering criteria, enabling precise alignment with adaptation goals.
        
        \item \textbf{Event Subscription}: enables services or service components to declaratively register interest in specific events (\eg a billing service subscribing to \texttt{HighErrorRateEvent}).  
    
        \item \textbf{Event Observation} (Contextual Awareness): Implements strategies to detect specified events, such as periodic polling (\eg check CPU every 30 seconds) or on-demand triggers (\eg API failure webhook). This component ensures timely responses to gradual trends (\eg memory leaks) and sudden anomalies (\eg DDoS attacks).  
    \end{itemize}  
    
\end{itemize}  

The following sections detail how these components collectively enable granular, context-aware adaptations while adhering to microservices’ decentralized ethos. By decoupling data collection, event detection, and action execution, AdaptiFlow allows developers to specify and gradually extend the adaptation logic without making major architectural changes.  

\subsection*{Metrics Collectors: Perception of the Environment}  
The foundation of self-adaptation lies in accurate perception. AdaptiFlow’s \textit{Metrics Collectors} are designed to gather both infrastructure-level (\eg CPU, memory) and business-level metrics (\eg transaction success rates, cache hit ratios). This dual focus ensures adaptations account for technical constraints and domain-specific requirements. For example, a \texttt{LatencyCollector} might monitor API response times, while a \texttt{CartAbandonmentCollector} tracks user behavior. Collectors enable services to expose their context for adaptation purposes. Once the collectors have been defined, it's easy to support both pull mechanisms (\eg Prometheus) and push mechanisms (\eg REST webhooks), accommodating diverse monitoring ecosystems. By decoupling data collection from analysis, AdaptiFlow allows developers to incrementally instrument services without overhauling existing systems.

\subsection*{Adaptation Actions: Executing Change}  
Adaptation actions translate decisions into runtime changes. AdaptiFlow enables the specification of adaptation actions that impact the:  
\begin{itemize}  
    \item \textbf{Infrastructure-Level}: Platform-specific or DevOps operations like \texttt{ScaleService} or \texttt{Re\-start\- a Container}.  
    \item \textbf{Business-Level}: Domain-specific adjustments such as \texttt{EnableFallbackRecommender} (switching to a lightweight algorithm) or \texttt{EnableCache}.  
\end{itemize}  

Action execution can either be synchronous (\eg immediate circuit breaking) or asynchronous (\eg batched log cleanup). Using our provided mechanism for adaptation logic specification, developers bind actions to events subscribers (services or service components). Each subscriber defines a list of adaptation actions that will be executed when the conditions defined in both the event and the subscriber are met. For example, a billing service might subscribe to \texttt{HighErrorRateEvent} to disable premium features temporarily, while a load balancer scales instances for the same event. 

\subsection*{Conditional Evaluators: Adaptive Decision-Making}  
Conditional Evaluators determine when events should trigger and also when subscribers should be notified to execute adaptation. AdaptiFlow implicitly provides two evaluator types: (1) \textbf{Threshold-Based Evaluators}, simple rules like \textit{GreaterThan} or \textit{Between}, ideal for tests on numerical data types (\eg scaling when CPU > 85\%) and (2) \textbf{Context-Aware Evaluators}, custom logic combining multiple metrics. For instance, a  \textit{PeakHourEvaluator} might disable non-essential features during high traffic only if cloud credits are low. Conditional Evaluators act like filters to know when adaptation actions will be executed.  

\subsection*{Event Specification}  
\label{subsec:event-spec}  

Events in AdaptiFlow are defined as logical combinations of conditions evaluated against collected metrics. The \textit{Event Specification} component allows developers to declaratively construct events using threshold-based or custom logic. Events serve as the bridge between raw metrics and actionable insights. here are some examples of events: (1) \textbf{Threshold-Based Events}, simple rules like \textit{HighCPUTEvent} (CPU > 80\%) or \textit{LowDiskSpaceEvent} (FreeDisk < 10\%) and (2) \textbf{Custom Events}, multi-condition rules such as \texttt{ServiceDegradationEvent} (latency > 1s \&\& error rate > 10\%).

\subsection*{Event Observation}  
\label{subsec:event-obs}  

The \textit{Event Observation} component implements strategies to detect specified events, balancing timeliness and resource efficiency:  (1) \textbf{Periodic Polling}, checks conditions at fixed intervals (\eg CPU every 30 seconds) for gradual trends like memory leaks and (2) \textbf{On-Demand Triggers}, event-driven checks (\eg during API failures) for rapid response to anomalies.
The \texttt{Observation Scheduler} orchestrates these strategies and developers can customize polling intervals or define their own observation logic, ensuring flexibility across scenarios. 

\subsection*{Event Subscription}  
\label{subsec:event-sub}  

AdaptiFlow’s \textit{Event Subscription} model allows services or their components to declaratively register interest in specific events. When an event triggers, the notification of subscribers consists of the execution of their adaptation actions. The adaptation action can be a local action inside the current service (\eg EnableCache, LowPowerMode), an API call to another service to execute some adaptation actions remotely (\eg OpenCirCuitBreaker) or the specification of another adaptation scenario (\eg DDoS Attack Mitigation).

This model supports hybrid architectures: a service can act as both a subscriber (\eg Auth service responding to \texttt{DDoSAttackEvent}) and an event emitter (\eg emitting \texttt{HighLatencyEvent}). Subscriptions can be dynamically updatable, allowing runtime adjustments without service restarts.

The preceding sections detailed AdaptiFlow’s core components, which collectively enable decentralized, context-aware adaptations. However, realizing these capabilities requires a systematic methodology to translate high-level adaptation requirements (\eg "prevent service outages during traffic spikes") into 

concrete implementations. This methodology\mdash the \textbf{Workflow for Enabling Adaptability} (Fig.~\ref{fig:adaptiflow})\mdash guides developers through six stages, each leveraging AdaptiFlow’s components to incrementally build self-adaptive logic:  

\begin{itemize}  
    \item \textbf{Metrics Collectors} operationalize the \textit{Identify Observables} stage, defining what data to gather.
    \item \textbf{Adaptation Actions} and \textbf{Event Subscription}  drive the \textit{Specify Event Subscribers} stage, linking events to executable responses.  
    \item \textbf{Conditional Evaluators} and \textbf{Event Specification} underpin the \textit{Specify Conditional Evaluators} and \textit{Specify Events} stages, mapping raw data to actionable triggers.  
    \item \textbf{Event Observation} implements the \textit{Configure Event Observation} stage, ensuring timely detection.  
\end{itemize}  

The following subsection elucidates this workflow, demonstrating how developers progress from abstract requirements (\eg mitigating DDoS attacks) to deployable adaptation logic.

\subsection*{Workflow for Enabling Adaptability with AdaptiFlow}  
\label{subsec:workflow}  

\begin{figure}[t]  
\centering  
\includegraphics[width=1\textwidth]{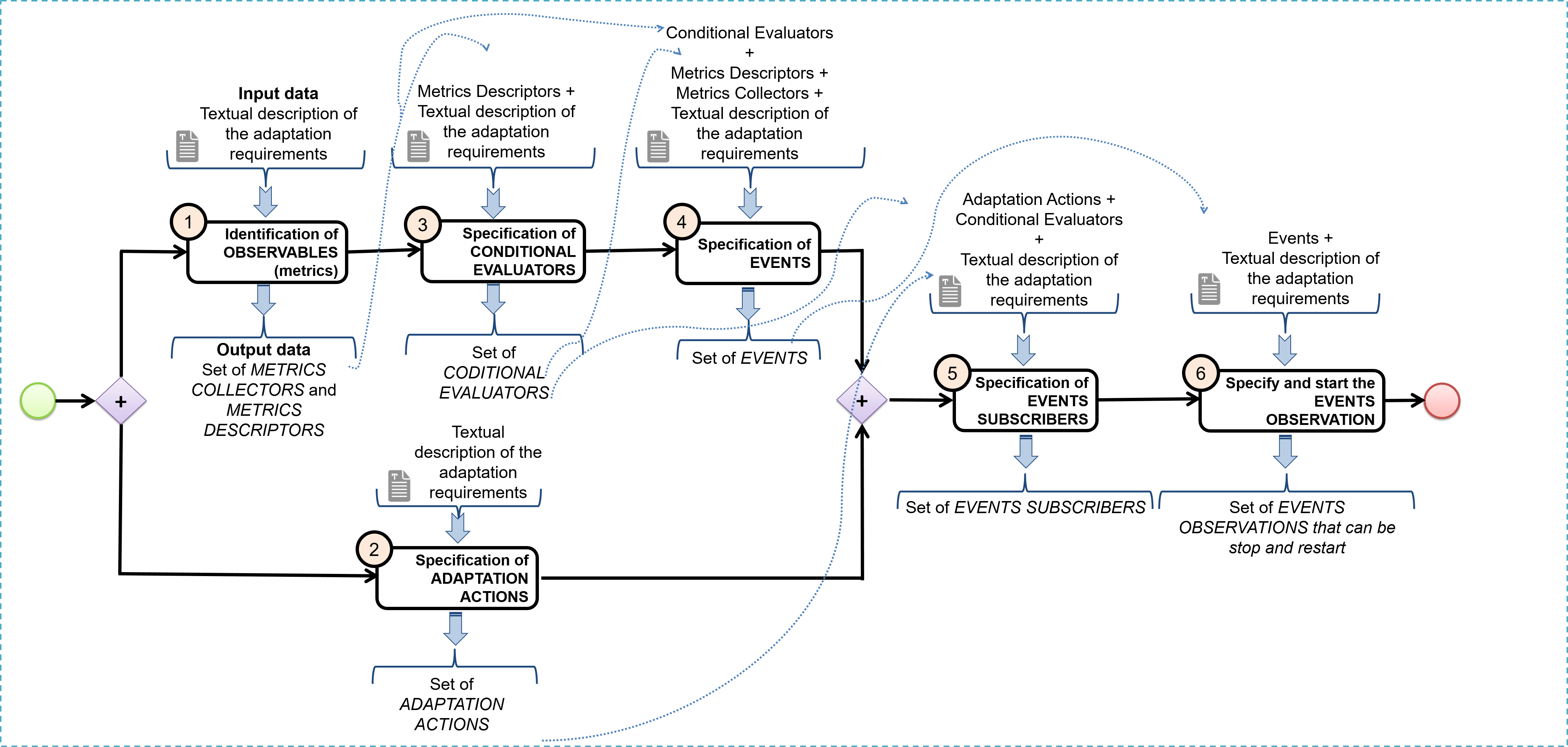}  
\caption{AdaptiFlow Workflow to enable adaptability in a given cloud microservices system using the textual description of the adaptation requirements or adaptation scenario description.}  
\label{fig:adaptiflow}  
\end{figure}

AdaptiFlow’s workflow (Fig.~\ref{fig:adaptiflow}) transforms textual adaptation requirements into executable adaptation logic through six systematic steps. To illustrate this process, consider a scenario where the TeaStore Persistence service \cite{von2018teastore} detects database timeouts and coordinates graceful degradation across dependent services (Auth, Recommender, Image, WebUI). The workflow begins with the identification of critical data points and culminates in the implementation of observation strategies, ensuring end-to-end adaptability.  

\textbf{Step 1\mdash Identify Observables:}  
The process starts by defining \textit{observables}—data sources required to monitor the system’s state. Developers parse adaptation requirements to determine which metrics (\eg database health, API latency) must be collected. For the Adaptable TeaStore scenario \cite{bliudze2024adaptable}, this involves creating \texttt{DatabaseHealthCollector} to track connection timeouts. AdaptiFlow provides abstract interfaces (\eg \texttt{IMetricsCollector}) to model these observables, decoupling data collection from downstream logic. Metrics descriptors define the structure of collected data (\eg sampling frequency, data type), enabling consistent interpretation across services.  

\textbf{Step 2\mdash Define Adaptation Actions (Parallel Step):}
Concurrently, developers specify \textit{adaptation actions}—concrete steps to achieve adaptation goals. These actions fall into two categories: \textit{infrastructure-level} (\eg restarting pods, scaling instances) and \textit{business-level} (\eg disabling recommendations, enabling maintenance modes). For Adaptable TeaStore, the Persistence service defines \texttt{EnableCache}, while the WebUI implements \texttt{EnableMaintenanceMode} and the recommender implements \texttt{Low\-Power\-Mode}. AdaptiFlow abstracts action execution through interfaces (\eg \texttt{IAdaptationAction}), allowing delegation to external tools (\eg Docker API, Kubernetes API) or custom logic. This separation ensures developers focus on defining \textit{what} to adapt, not \textit{how} to implement low-level operations.

Steps 1 and 2 constitute the \textit{preparation phase}, equipping services with the interfaces required for autonomic management. By defining metrics collectors (Step 1), services expose monitoring endpoints that provide real-time insights into their state (\eg database health, API latency). Simultaneously, specifying adaptation actions (Step 2) establishes control points—actuators that enable runtime adjustments (\eg restarting pods, toggling features). This aligns with the Autonomic Computing Vision (ACV), where "the managed element is adapted to enable the autonomic manager to monitor and control it" \cite{kephart2003vision}. AdaptiFlow operationalizes this by decoupling data collection and action execution into reusable interfaces (\texttt{IMetricsCollector}, \texttt{IAdaptationAction}), effectively transforming non-adaptive services into autonomic elements. Once prepared, services offer standardized APIs for observation and control, enabling the autonomic manager to implement adaptation scenarios.  

\textbf{Step 3\mdash Specify Conditional Evaluators:}  
Conditional evaluators encode the logic for triggering events and filtering subscribers. This step involves dual evaluations: (1) \textit{event evaluators} determine if an event should trigger (\eg \texttt{DatabaseTimeoutEvaluator} checks for consecutive timeouts), and (2) \textit{subscriber evaluators} decide which subscribers should act (\eg Auth service acts only after a 5-minute outage). Evaluators leverage collected metrics and can integrate external APIs (\eg machine learning models for anomaly detection). For our Adaptable TeaStore, an \texttt{UnHealthyDatabaseEvaluator} combines database health metrics with service dependency statuses to assess system stability.  

\textbf{Step 4\mdash Specify Events:}  
Events semantically encapsulate adaptation scenarios. Developers bind evaluators and metrics to named events (\eg \texttt{DatabaseUnavailableEvent}), which act as triggers for non-coordinated actions. Events inherit from AdaptiFlow’s \texttt{ConditionalEvent} base class, enabling reuse across scenarios. For Adaptable TeaStore, the \texttt{DatabaseUnavailableEvent} is defined using the \texttt{LocalDatabaseMetricsCollector} and \texttt{UnHealthyDatabaseEvaluator}, ensuring it triggers only when timeout thresholds are breached. Events provide human-readable context (\eg “database\_unavailable”), aligning with adaptation goals described in requirements.  

\textbf{Step 5\mdash Specify Event Subscribers:}  
Subscribers declaratively register for events and define action execution strategies. A subscriber comprises (1) a list of adaptation actions and (2) a conditional evaluator to filter notifications. In Adaptable TeaStore, the Persistence service subscribes to \texttt{DatabaseUnavailableEvent} with an \texttt{EnableCache} action, while the WebUI service switches to maintenance mode. Subscribers can be granular (\eg specific UI components) or service-wide. AdaptiFlow supports strategies like \textit{immediate execution} (act on first trigger) or \textit{event counting} (act after N occurrences), offering flexibility akin to MAPE-K’s planning phase without fully implementing a planner.  

\textbf{Step 6\mdash Configure Event Observation:}  
The final step defines \textit{how} events are detected. AdaptiFlow supports periodic polling (\eg check database health every 10s) or on-demand triggers (\eg during API failures). For Adaptable TeaStore, the \texttt{ObservationScheduler} uses periodic checks for database health. Developers can extend the \texttt{AbstractObservationScheduler} class to implement custom strategies (\eg event-driven checks via message queues), ensuring adaptability to platform constraints. Observation configurations are decoupled from event logic, allowing runtime adjustments without disrupting active adaptations.  

This structured workflow ensures systematic implementation of adaptation scenarios while preserving microservices’ autonomy. Section~\ref{secn:casestudy} validates AdaptiFlow’s efficacy through three scenarios in the Adaptable TeaStore: \textit{self-healing} (database recovery), \textit{self-protection} (DDoS mitigation), and \textit{self-optimization} (traffic management).

\section{Case Study: Building the Adaptable TeaStore}
\label{secn:casestudy}

\begin{figure}[p]  
\centering  
\includegraphics[angle=90,height=0.95\textheight]{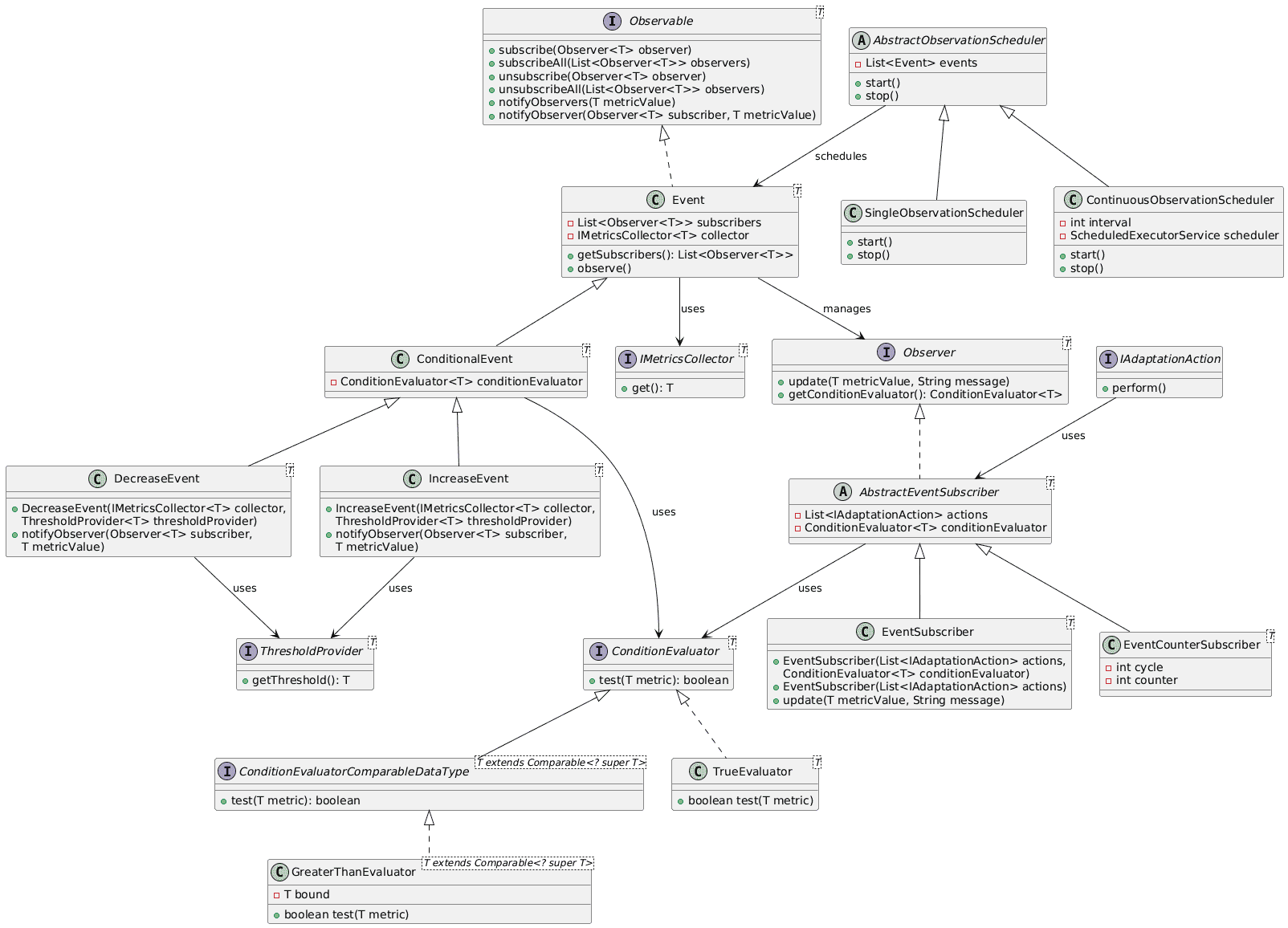}
\caption{AdaptiFlow class diagram.}
\label{fig:class-diagram}  
\end{figure} 

Validation of AdaptiFlow's design principles was carried out through comprehensive experiments with the TeaStore benchmark in order to provide an adaptable version: Adaptable TeaStore previously described by Bliudze et al \cite{bliudze2024adaptable}. This case study has two main objectives: (1) to demonstrate the practical implementation of our abstraction layers (Figure~\ref{fig:class-diagram}) and (2) to evaluate the effectiveness of the framework through distinct adaptation scenarios. We chose TeaStore \cite{von2018teastore} for its representative cloud-native architecture comprising five interdependent services (Auth, Persistence, Recommender, Image, and WebUI). The experimental methodology systematically examines three autonomous capabilities by implementing three adaptation scenarios: self-healing (service recovery), self-protection (DDoS mitigation), and self-optimization (traffic management).

\subsection{Experimental Setup}  
\label{subsec:setup}  

The experiments were conducted on a Docker-based \cite{merkel2014docker} environment with Portainer CE \cite{portainerWelcomePortainer} for container management. Each TeaStore service (Auth, Persistence, Recommender, Image, WebUI) was instrumented with AdaptiFlow’s abstraction layer, exposing standardized REST APIs and Java classes for metrics collection and adaptation action execution. We use object-oriented subclassing as the instrumentation method. The AdaptiFlow framework library (compiled with JDK 11) provides base classes and interfaces for implementing metrics collectors, adaptation actions and event handlers.

We utilized the HTTP load generator \cite{von2018teastore} with Limbo \cite{KiHeKo2014-ICPEDemo-LIMBO} for modeling load intensities just like it is done in the original TeaStore. We containerized the two components of the load generator (the director and the load generator) for Docker compatibility. Load intensity was controlled via three CSV profiles: (1) \textit{increasingLowIntensity.csv} for gradual ramp-up, (2) \textit{increasingMedIntensity.csv} for moderate ramp-up and (3) \textit{increasingHighIntensity.csv} for aggressive ramp-up.

For simplicity, we used the \textit{increasingHighIntensity} profile to simulate DDoS attack conditions, while \textit{increasingMedIntensity} tested self-optimization thresholds. Locust was present in the original configuration, but we intentionally ignored it as the Limbo HTTP load generator was sufficient for our experiments.

Each service’s Docker container included: (1) \textit{Metrics Collectors} (Infrastructure / Business-level), (2) \textit{Adaptation Actions} (Business-level only) and \textit{Event Handlers}. As our main objective was to define the abstraction layers needed to specify metrics collectors and adaptation actions, we have not addressed the implementation details of adaptation actions at the infrastructure level (\eg stopping or restarting containers), since applications such as \texttt{Portainer} \cite{portainerWelcomePortainer} demonstrate the feasibility of such actions in Docker and Kubernetes environments. We focused more on implementing adaptation actions linked to the business logic of the various microservices (\eg optimizing recommendations, enabling/disabling caching, using an external provider for images). In addition, the validation focused on Docker; Kubernetes behavior was verified through API responses, but was not tested in cluster orchestration scenarios.

This setup enabled systematic evaluation of AdaptiFlow’s ability to translate adaptation requirements into runtime behavior adjustments. The primary result of our experimental validation is the successful implementation of three distinct adaptation scenarios using the AdaptiFlow abstraction layer. The key success metric was functional: for each scenario, the system correctly detected the triggering condition and executed the predefined adaptation actions across the affected services without failure in repeated trials.

The following subsections detail our implementation of three autonomic scenarios, demonstrating how AdaptiFlow’s abstraction layer (Figure~\ref{fig:class-diagram}) bridges the gap between non-adaptive services and self-managing elements.

\subsection{Self-Healing: Database Unavailability}
\label{subsec:self-healing}

\begin{figure}[p]  
\centering  
\includegraphics[angle=90,height=0.95\textheight]{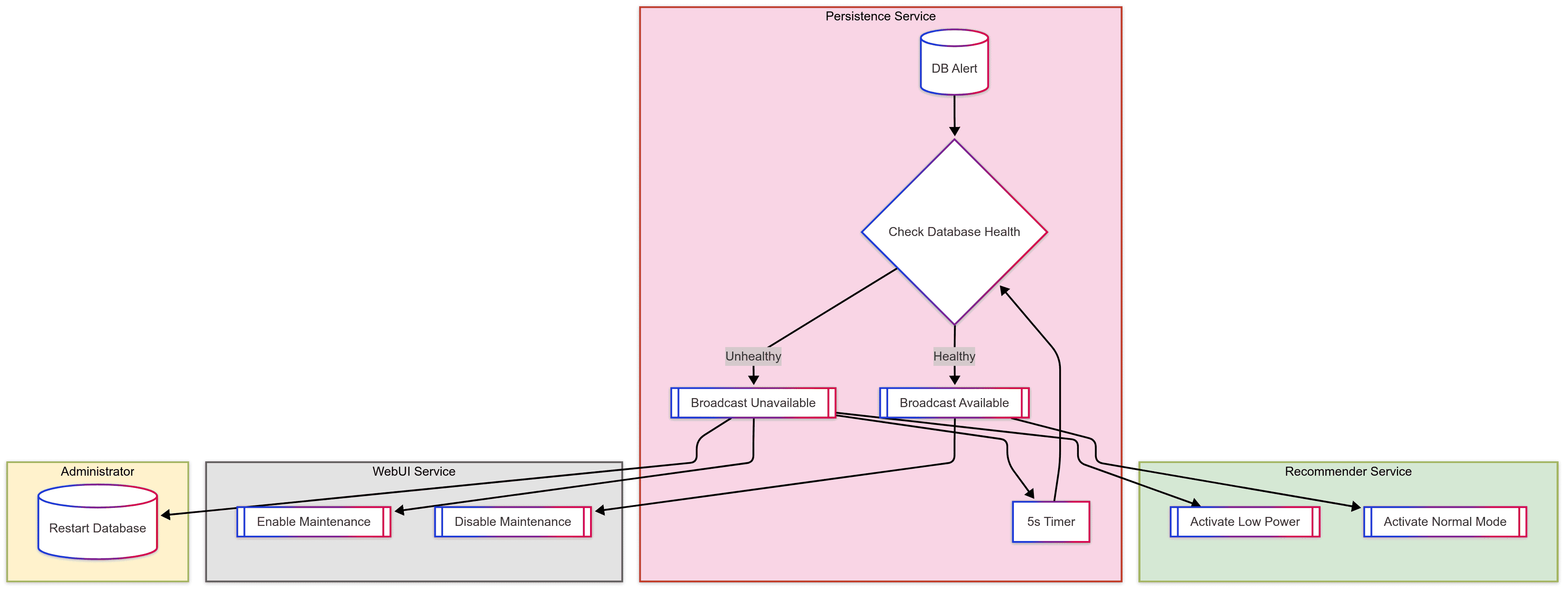}
\caption{The database unavailable mitigation flow diagram}  
\label{fig:database-unavailable-flow}  
\end{figure} 

\begin{table}[t]
\tabsize
\centering
\caption{Self-healing scenario implementation mapping to AdaptiFlow workflow}
\label{tab:self-healing-workflow}
\settowidth{\tabwidthi}{\tabsize Specify Conditional Evaluators}
\settowidth{\tabwidthii}{\tabsize \textit{DatabaseUnavailableEventBroadcast,}}
\settowidth{\tabwidthiii}{\tabsize Tracks request rates, IP patterns, and error frequencies}
\setlength{\tabwidthiii}{\textwidth-\tabwidthi-\tabwidthii-4\tabcolsep}
\begin{tabular}{@{}
    >{\raggedright}p{\tabwidthi}
    >{\raggedright}p{\tabwidthii}
    >{\raggedright}p{\tabwidthiii}
    @{}
}
\toprule
\textbf{Workflow Step} & \textbf{Key Classes} & \textbf{Implementation Purpose}
\tabularnewline\toprule
Identify Observables & \textit{LocalDatabaseMetricsCollector} & Gathers database health metrics via JDBC checks
\tabularnewline\midrule
Define Adaptation Actions & \textit{DatabaseAvailableEventBroadcast, DatabaseUnavailableEventBroadcast}, \textit{EnableMaintenanceMode, DisableMaintenanceMode}, \textit{EnableCache, DisableCache}, \textit{LowPowerMode, NormalMode} & Declares business-level actuators
\tabularnewline\midrule
Specify Conditional Evaluators & \textit{HealthyDatabaseEvaluator, UnHealthyDatabaseEvaluator} & Encapsulates detection logic of the database health status
\tabularnewline\midrule
Specify Events & \textit{DatabaseAvailableEvent, DatabaseUnavailableEvent} & Links metrics to semantic adaptation triggers
\tabularnewline\midrule
Specify Event Subscribers & \textit{EventSubscriber} & Registers service-specific response actions
\tabularnewline\midrule
Configure Event Observation & \textit{ContinuousObservationScheduler} & Implements periodic detection strategy
\tabularnewline\bottomrule
\end{tabular}
\end{table}


\paragraph{Scenario overview (Figure~\ref{fig:database-unavailable-flow}):} The system is deployed in a barebone configuration with local services (Auth, Recommender, Image, Persistence, WebUI). The Persistence service detects timeouts from the local database due to an unexpected interruption. It triggers adaptation actions across dependent services (Auth, Recommender, Image, WebUI) to gracefully degrade functionality. The WebUI displays a maintenance message, and the system administrator is alerted to restart the database. Once restored, services resume normal operation.

In this scenario, the Persistence microservice serves as the central adaptation coordinator, monitoring database health through three key metrics: \textit{connection status} (boolean), \textit{query response times} (milliseconds), and \textit{active connection counts}. Upon detecting failures, it triggers cascading adaptations across dependent components. The WebUI responds by displaying maintenance pages, while the Recommender service dynamically adjusts its algorithm between normal operation (popular items only) and low-power mode (no recommendations).

\paragraph{Implementation Methodology:}
The implementation follows AdaptiFlow's six-step workflow for enabling adaptability (Section~\ref{subsec:workflow}), systematically translating requirements into executable adaptation logic:

\textbf{Step 1\mdash Identify Observables:} 
The \textit{LocalDatabaseMetricsCollector} class implements the \textit{IMetricsCollector} interface to monitor four critical database health indicators: response times, network status, active connections, and pending queries. These observables provide real-time insights into database health through JDBC health checks and connection pool monitoring.

\textbf{Step 2\mdash Define Adaptation Actions:} 
Business-level actuators were implemented across services:
\begin{itemize}
    \item \textit{DatabaseAvailableEventBroadcast} / \textit{DatabaseUnavailableEventBroadcast}, \textit{EnableCache} / \textit{DisableCache} in Persistence service
    \item \textit{EnableMaintenanceMode} / \textit{DisableMaintenanceMode} in WebUI
    \item \textit{LowPowerMode} / \textit{NormalMode} in Recommender
\end{itemize}
Four adaptation patterns coordinate the system response. \textit{Event broadcasting} in persistence service propagates status changes via REST notifications, while cache management dynamically \textit{enables/disables} caching to improve system fault tolerance. The UI degradation pattern activates maintenance displays, and service throttling reduces computational load through the Recommender's power modes. These actions demonstrate AdaptiFlow's ability to easily combine infrastructure and business-level adaptations.

\textbf{Step 3\mdash Specify Conditional Evaluators:} 
The system employs two condition evaluators with distinct triggering mechanisms. The \textit{UnHealthyDatabaseEvaluator} activates when response times exceed 5000ms or when the network status deviates from expectations. Conversely, the \textit{HealthyDatabaseEvaluator} requires both sub-5000ms response times and proper network status before signaling recovery:
\begin{itemize}
    \item \texttt{UnHealthyDatabaseEvaluator} triggers when response times exceed 5000ms or network status deviates
    \item \texttt{HealthyDatabaseEvaluator} requires both sub-5000ms response times and proper network status
\end{itemize}
These implement the \texttt{ConditionEvaluator} interface with custom validation logic.

\textbf{Step 4\mdash Specify Events:} 
The \textit{DatabaseUnavailableEvent} and \textit{DatabaseAvailableEvent} extend \textit{ConditionalEvent}, combining the metrics collector with their respective evaluators. This event abstraction serves as the semantic bridge between raw metrics and adaptation triggers.

\textbf{Step 5\mdash Specify Event Subscribers:} 
Services register interest through \texttt{EventSubscriber} instances. In this scenario, only the persistence service subscribes to database availability and unavailability events and then notifies the other services (WebUI and Recommender) so that they can execute their adaptation actions.

\textbf{Step 6\mdash Configure Event Observation:} 
A \texttt{ContinuousObservationScheduler} with 5-second polling intervals monitors events. The scheduler initiates the adaptation cycle by periodically invoking metrics collection and evaluation.

\paragraph{Implementation Notes:} Current limitations include REST-based event propagation (planned upgrade to message brokers) and manual database recovery (future automation target). These constraints were intentionally maintained during validation to isolate and demonstrate the framework's core capabilities.

\subsection{Self-Protection: Mitigating DDoS Attacks}

\begin{figure}[p]
\centering
\includegraphics[angle=-90,origin=c,height=0.75\textheight]{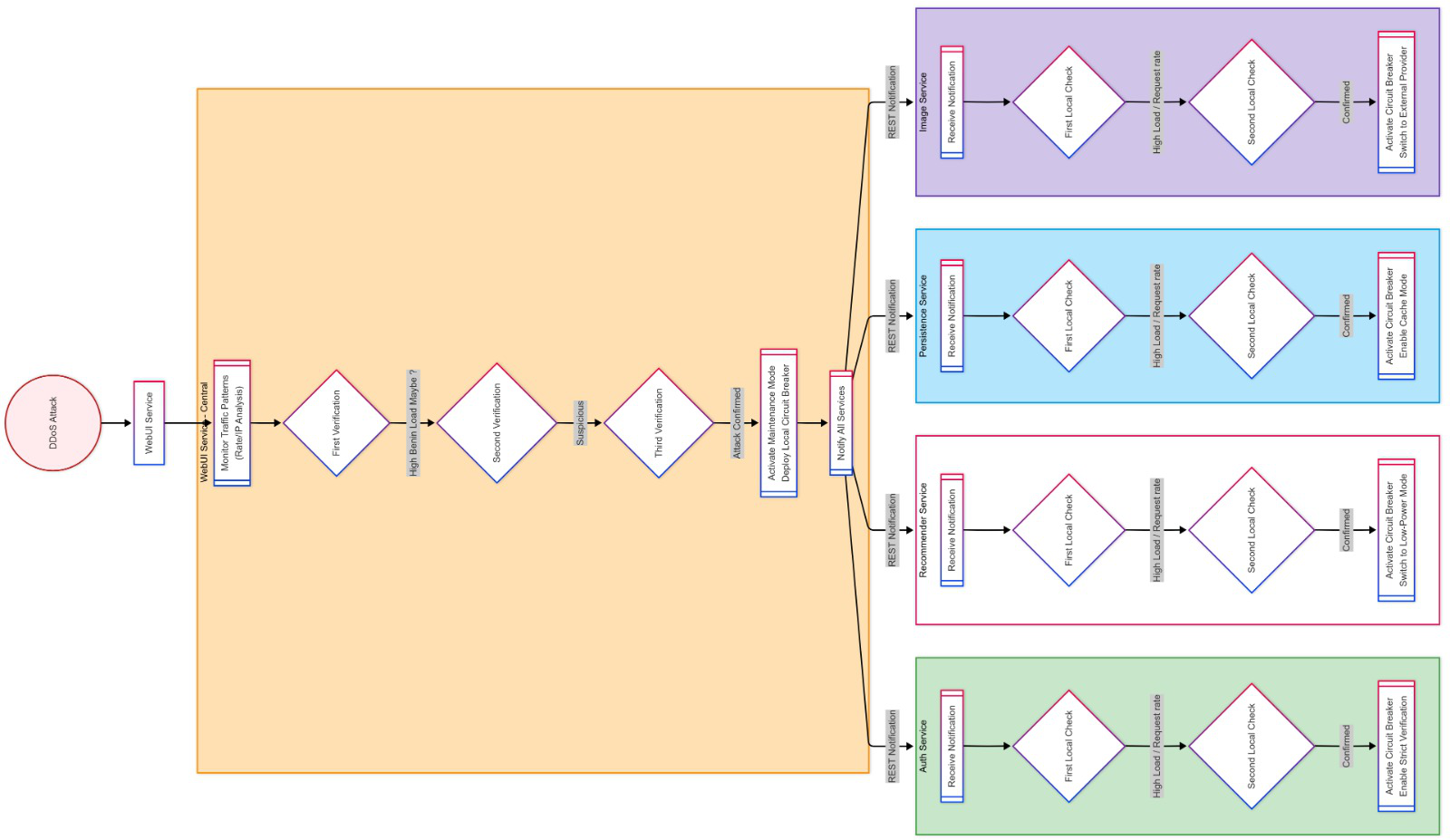}
\caption{Malicious traffic adaptation workflow}
\label{fig:malicious-flow}
\end{figure}

\begin{table}[t]
\tabsize
\centering
\caption{DDoS mitigation scenario implementation mapping to AdaptiFlow workflow}
\label{tab:ddos-workflow}
\settowidth{\tabwidthi}{\tabsize Specify Conditional Evaluators}
\settowidth{\tabwidthii}{\tabsize \textit{ContinuousObservationScheduler}}
\settowidth{\tabwidthiii}{\tabsize Tracks request rates, IP patterns, error frequencies}
\setlength{\tabwidthii}{\textwidth-\tabwidthi-\tabwidthiii-4\tabcolsep}
\begin{tabular}{@{}
    >{\raggedright}p{\tabwidthi}
    >{\raggedright}p{\tabwidthii}
    p{\tabwidthiii}
    @{}
}
\toprule
\textbf{Workflow Step} & \textbf{Key Classes} & \textbf{Implementation Purpose}
\tabularnewline\toprule
Identify Observables & \textit{LocalRequestMetricsCollector} & Tracks request rates, IP patterns, error frequencies
\tabularnewline\midrule
Define Adaptation Actions & \textit{DDoSAttackEventBroadcast}, \textit{EnableMaintenanceMode}, \textit{OpenCircuitBreaker}, \textit{CloseCircuitBreaker}, \textit{LowPowerMode}, \textit{EnableExternalImageProvider} & Declares protection actuators across services
\tabularnewline\midrule
Specify Conditional Evaluators & \textit{DDoSEvaluator}, \textit{NonDDoSEvaluator} & Implements threshold verification logic
\tabularnewline\midrule
Specify Events & \textit{MaliciousTrafficEvent, BenignTrafficEvent} & Defines attack detection triggers
\tabularnewline\midrule
Specify Event Subscribers & \textit{EventCounterSubscriber} & Manages multi-stage verification requirements
\tabularnewline\midrule
Configure Event Observation & \textit{ContinuousObservationScheduler} & Coordinates periodic and event-driven detection
\tabularnewline\bottomrule
\end{tabular}
\end{table}

\paragraph{Scenario overview (Figure~\ref{fig:malicious-flow}):} The system responds to a \textit{Distributed Denial-of-Service} attack through a layered verification and adaptation protocol. The WebUI service acts as the primary detector, analyzing traffic patterns through request rate monitoring before declaring an attack only after three consecutive verification stages. Upon confirmation, it simultaneously activates the local circuit breaker and broadcasts REST notifications to all dependent services. Each recipient service—Auth, Recommender, Persistence, and Image—performs its own dual-layer validation of attack indicators before implementing specialized protections. First of all, the entire services activate its local circuit breaker, then the Recommender downgrades to low-power operation mode. The Persistence service shifts to cache-only operations, and the Image service reroutes requests to external providers. 

This scenario describes a multi-stage approach that combines centralized detection with decentralized execution, ensuring attack mitigation while preserving critical functionality through coordinated circuit breaker deployment and service-specific degradation modes. The entire process emphasizes verification rigor, requiring five total confirmation checks (three central and two local) before full protective measures engage, preventing false positives during normal traffic spikes.

\paragraph{Implementation Methodology:}
The implementation follows AdaptiFlow's six-step workflow:

\textbf{Step 1\mdash Identify Observables:} 
In this scenario, we employ uniform metric collection across all services through \textit{LocalRequestMetricsCollectors}, monitoring \textit{request rates}, \textit{IP patterns}, \textit{error frequencies} and \textit{requests details}. The WebUI service initiates detection using threshold-based evaluators (\textit{DDoSEvaluator} / \textit{NonDDoSEvaluator}) that analyze requests per second over 60-second windows: more complex verifications could have been done, but we chose to simplify the scenario for our experiments. While all services can track contextual traffic metrics, at the beginning of the process, only WebUI detects and triggers system-wide adaptations when its evaluator detects sustained rates exceeding 300 requests/second. Then, WebUI notifies all services to start their own decentralized monitoring and detection.

\textbf{Step 2\mdash Define Adaptation Actions:} 
Adaptations cascade through a two-phase protocol. The WebUI service first activates maintenance pages and the circuit breaker after triple-confirmed detection, then broadcasts alerts via REST. Recipient services implement specialized protections: Auth activates circuit breaker, Recommender switches to low power mode (no recommendations), Persistence prioritizes cached data, and Image services reroute to external providers. Each service independently verifies attack conditions through dual local checks before executing actions, preventing single-point failures.

\begin{itemize}
\item WebUI: \textit{EnableMaintenanceMode, DisableMaintenanceMode}, \textit{OpenCircuitBreaker}, \textit{CloseCircuitBreaker}, and \textit{DDoSAttackEventBroadcast}

\item Auth: \textit{OpenCircuitBreaker} and \textit{CloseCircuitBreaker}

\item Recommender: \textit{LowPowerMode}, \textit{NormalMode}, \textit{OpenCircuitBreaker} and \textit{CloseCircuitBreaker}

\item Image: \textit{EnableExternalImageProvider}, \textit{DisableExternalImageProvider}, \textit{OpenCircuitBreaker} and \textit{CloseCircuitBreaker}
\end{itemize}
These actions demonstrate AdaptiFlow's support for heterogeneous response strategies across services.

\textbf{Step 3\mdash Specify Conditional Evaluators:} 
\textit{DDoSEvaluator} implements threshold-based attack detection, triggering when request rates exceed 300 req/sec. The complementary \textit{NonDDoSEvaluator} handles recovery conditions.

\textbf{Step 4\mdash Specify Events:} 
\textit{MaliciousTrafficEvent} and \textit{BenignTrafficEvent} extend \textit{ConditionalEvent}, combining metrics collectors with evaluators. These event abstractions encapsulate attack semantics while decoupling detection from response logic.

\textbf{Step 5\mdash Specify Event Subscribers:} 
WebUI uses \textit{EventCounterSubscriber}, requiring three consecutive threshold breaches before triggering adaptations, reducing false positives. Downstream services implement dual-verification subscribers that perform two local confirmations before executing protections. This hierarchical subscription model balances system-wide awareness with local context validation.

\textbf{Step 6\mdash Configure Event Observation:} 
A \texttt{ContinuousObservationScheduler} with 5-second polling intervals monitors events across all services. The scheduler initiates the adaptation cycle by periodically invoking metrics collection and evaluation.

\paragraph{Technical Insights:} The event-driven architecture uses the \textit{ConditionalEvent} wrapper around metric streams, with WebUI’s \textit{EventCounterSubscriber} requiring three threshold breaches before triggering adaptations. Recipient services reuse the same evaluator classes but configure two-step verification. REST notifications propagate through standardized adaptation endpoints, enabling heterogeneous action execution—from UI changes (\textit{EnableMaintenanceMode}) to infrastructure adjustments (\textit{EnableExternalImageProvider}). The code structure demonstrates AdaptiFlow’s template pattern, where services implement shared interfaces (\textit{IAdaptationAction}, \textit{ConditionEvaluator}) while customizing verification thresholds and action combinations. REST notifications propagate through standardized endpoints, though future versions will also integrate message brokers for enhanced reliability.

\subsection{Self-Optimization: Handling Benign Traffic Surges}

\begin{figure}[p]
\centering
\includegraphics[angle=90,height=0.94\textheight]{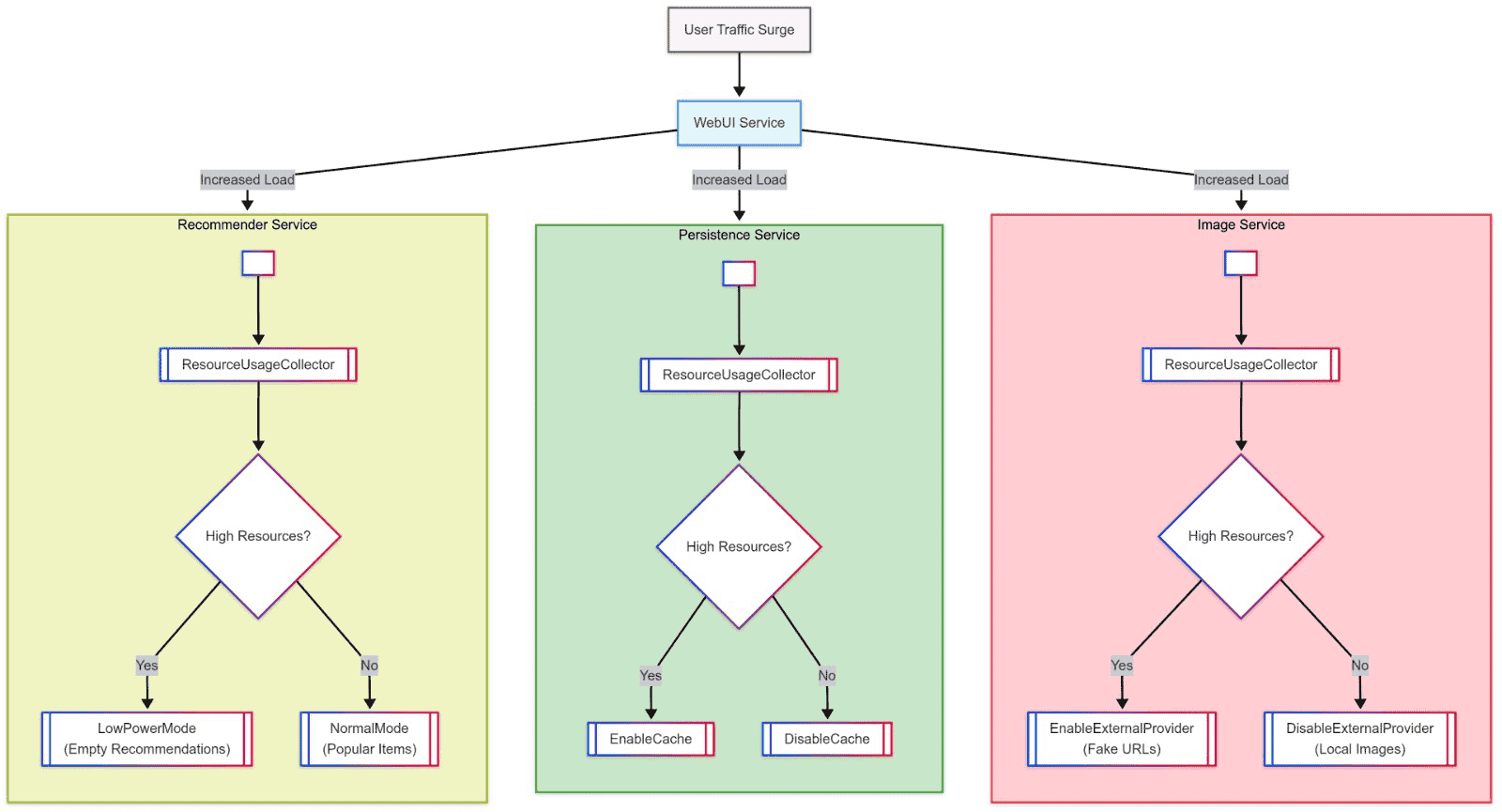}
\caption{Benign traffic adaptation workflow}
\label{fig:benign-flow}
\end{figure}

\begin{table}[t]
\tabsize
\centering
\caption{Self-optimization scenario implementation mapping to AdaptiFlow workflow}
\label{tab:optimization-workflow}
\settowidth{\tabwidthi}{\tabsize Specify Conditional Evaluators}
\settowidth{\tabwidthii}{\tabsize \textit{DecreaseResourceUsageEvaluator}}
\settowidth{\tabwidthiii}{\tabsize Monitors CPU/memory utilization across services}
\setlength{\tabwidthiii}{\textwidth-\tabwidthi-\tabwidthii-4\tabcolsep}
\begin{tabular}{@{}
    >{\raggedright}p{\tabwidthi}
    >{\raggedright}p{\tabwidthii}
    >{\raggedright}p{\tabwidthiii}
    @{}
}
\toprule
\textbf{Workflow Step} & \textbf{Key Classes} & \textbf{Implementation Purpose} 
\tabularnewline\toprule
Identify Observables & \textit{ResourceUsageCollector} & Monitors CPU/memory utilization across services
\tabularnewline\midrule
Define Adaptation Actions & \textit{LowPowerMode, EnableCache, NormalPowerMode, Disable, EnableExternalImageProvider} & Implements service-specific optimization strategies
\tabularnewline\midrule
Specify Conditional Evaluators & \textit{IncreaseResourceUsageEvaluator}, \textit{DecreaseResourceUsageEvaluator} & Defines threshold-based optimization triggers
\tabularnewline\midrule
Specify Events & \textit{TrafficIncreaseEvent}, \textit{TrafficDecreaseEvent} & Encapsulates resource-based adaptation triggers
\tabularnewline\midrule
Specify Event Subscribers & \textit{EventSubscriber} & Registers autonomous optimization responses
\tabularnewline\midrule
Configure Event Observation & \textit{ContinuousObservationScheduler} & Implements periodic resource monitoring
\tabularnewline\bottomrule
\end{tabular}
\end{table}

\paragraph{Scenario overview (Figure~\ref{fig:benign-flow}):} The system autonomously adapts to legitimate traffic spikes through context-aware resource optimization when infrastructure scaling is unavailable. Faced with surging user demand, the WebUI service triggers decentralized adaptations across dependent components without centralized coordination. The Recommender service switches to no recommendations, reducing computational load. Simultaneously, the Persistence service activates caching mechanisms, prioritizing frequently accessed data to alleviate database pressure. Image processing workloads shift dynamically to an external provider. Each service independently monitors resource utilization through dedicated collectors—CPU and memory-to calibrate adaptations, enabling system-wide load balancing. 

In this scenario, we use an uncoordinated optimization strategy that demonstrates AdaptiFlow's ability, where autonomous decisions at the service level collectively contribute to stabilizing the system despite resource constraints. This stabilization is sometimes achieved at the expense of the user experience, depending on the adaptation actions chosen.

\paragraph{Implementation Methodology:}
The implementation follows AdaptiFlow's six-step workflow for decentralized optimization:

\textbf{Step 1\mdash Identify Observables:} 
we employ decentralized resource monitoring through service-specific \textit{ResourceUsageCollectors}, tracking CPU and memory utilization. Each service (Recommender, Persistence, Image) independently observes metrics via hash maps containing numeric percentages for \textit{cpu_usage} and \textit{memory_usage}. Threshold-based evaluators trigger adaptations when CPU exceeds 75\% or memory surpasses 80\%, with recovery initiated when both metrics drop below 60\%. This dual-threshold approach prevents oscillations during borderline resource conditions while maintaining granular control over optimization triggers.

\textbf{Step 2\mdash Define Adaptation Actions:} 
Services implement autonomous optimization strategies when thresholds breach. The Recommender reduces computational load by switching to no recommendations via \textit{LowPowerMode}, cutting recommendation logic overhead. Persistence activates caching mechanisms (\textit{EnableCache}) to minimize database queries, while the Image service offloads processing to an external provider through \textit{EnableExternalImageProvider}. Recovery actions revert systems to normal operations when resource usage stabilizes.

\begin{itemize}
    \item Recommender: \textit{LowPowerMode}, \textit{NormalMode}
    \item Persistence: \textit{EnableCache}, \textit{DisableCache}
    \item Image: \textit{EnableExternalImageProvider}, \textit{DisableExternalImageProvider}
\end{itemize}
These localized adaptations occur without inter-service coordination, allowing parallel optimization while preserving core functionality.

\textbf{Step 3\mdash Specify Conditional Evaluators:} 
The \textit{IncreaseResourceUsageEvaluator} triggers optimizations when either CPU or memory exceeds upper limits (default: 75\% CPU, 80\% memory), while \textit{DecreaseResourceUsageEvaluator} restores normal operations when both metrics fall below recovery thresholds (60\%). This asymmetric logic prioritizes rapid response to overloads while requiring sustained improvement for recovery.

\textbf{Step 4\mdash Specify Events:} 
The \textit{TrafficIncreaseEvent} and \textit{TrafficDecreaseEvent} extend \texttt{Con\-di\-tion\-al\-Event}, combining resource metrics with their respective evaluators. These events provide semantic context for optimization triggers while maintaining service isolation.

\textbf{Step 5\mdash Specify Event Subscribers:} 
Services use \textit{EventSubscriber} instances to register their optimization responses directly. The granular subscription model allows per-service threshold customization while maintaining standardized interfaces. For example, Image services can implement higher CPU thresholds (85\%) than Recommender services (75\%).

\textbf{Step 6\mdash Configure Event Observation:} 
A \texttt{ContinuousObservationScheduler} with configurable polling intervals (default: 5 seconds) monitors resource events. The scheduler initiates periodic resource checks, balancing detection responsiveness with system overhead.

\paragraph{Technical Insights:} Decentralized decision-making leverages identical threshold patterns across services with service-specific interpretations. The \textit{IncreaseResourceUsageEvaluator} class triggers optimizations when either CPU or memory exceeds upper limits, while \textit{DecreaseResourceUsageEvaluator} restores normal operations when both metrics fall below recovery thresholds. This asymmetric logic prioritizes rapid response to overloads while requiring sustained improvement for recovery, balancing stability with responsiveness. Services maintain isolation, Rescommender never influences Persistence caching decisions.

The architecture uses standardized \textit{ConditionalEvent} wrappers around resource metrics, enabling code reuse across services while allowing threshold customization. HashMaps structure metrics with strict key conventions (\textit{cpu_usage}, \textit{memory_usage}) for evaluator compatibility. Five-second polling intervals (configurable via \textit{EVENT_LISTENING_INTERVAL_MS}) balance detection speed with overhead. Though services share adaptation patterns, each can maintain separate threshold configurations.

\section{Conclusion and future work}
\label{secn:conclusion}

The development and evaluation of AdaptiFlow demonstrate a workflow and abstraction layer for integrating self-adaptive capabilities into cloud-native microservices. More precisely, AdaptiFlow addresses the critical challenge of enabling self-adaptation in cloud-native microservices through an abstraction layer focused on the \textbf{Monitor} and \textbf{Execute} phases of the MAPE-K loop. By decoupling metrics collection and action execution from adaptation logic, the framework provides standardized interfaces that transform conventional services into autonomic elements with minimal code changes. While our primary focus was on abstracting the monitoring and execution phases, we introduced lightweight rule-based mechanisms for the \textbf{Analyze} and \textbf{Plan} phases to validate the core architecture through realistic scenarios. This pragmatic approach demonstrates how decentralized, event-driven adaptations can emerge from localized decisions while maintaining system-wide coherence.  

The framework’s strength lies in its dual-purpose API design: developers instrument services with metrics collectors and actuators using unified interfaces, while adaptation stakeholders, such as human operators or AI-based agents, leverage these primitives to implement diverse control policies. Our validation through the \textbf{Adaptable TeaStore} case-study~\cite{bliudze2024adaptable} confirms AdaptiFlow’s practicality across three critical adaptation objectives: self-healing, self-protection, and self-optimization.

As part of the future work, we plan to focus on three key enhancements:  
\begin{itemize}  
    \item \textbf{Formal Coordination}: A current limitation is the potential for conflicting adaptations when multiple, independent scenarios are triggered within the same service. To address this, we plan to integrate a formal coordination engine, such as \textit{JavaBIP} \cite{JavaBIP-spe} or \textit{Multi-Bach} \cite{JACQUET2021102579}, to explicitly model the composition of concurrent adaptation behaviors. This would allow us to rigorously specify interaction protocols and prevent undesirable system states. 
    \item \textbf{Adaptation Language}: Exploration of a domain-specific language (DSL) for declarative rule or adaptation scenario specification, lowering the barrier for non-programmers to define adaptation policies.  
    \item \textbf{Intelligent Adaptation}: Implementation and evaluation of AI-driven strategies using the AWARE framework \cite{sanwouo2025breaking}, comparing machine learning approaches with traditional rule-based methods in terms of responsiveness and resource efficiency.  
\end{itemize}

By bridging the gap between autonomous computing theory and cloud computing practice, AdaptiFlow consolidates the foundations of self-adaptive systems, in which microservices autonomously navigate dynamic environments, while developers retain full (ideal case) or moderate (if necessary) control over adaptation semantics. This balance between automation and flexibility makes the framework an essential tool for resilient, efficient cloud architectures at a time of ever-increasing operational complexity.  


\nocite{*}
\bibliographystyle{eptcs}
\bibliography{generic}

\end{document}